\documentclass{iau} 
\usepackage{graphicx}

\title[IAUS 315.~~Molecule-Forming Clouds in the Galaxy] 
{A Global View of Molecule-Forming Clouds in the Galaxy}

\author[Gibson et al.]   
{
Steven~J. Gibson$^{1}$, 
Ward~S. Howard$^{1,2,3}$, 
Christian~S. Jolly$^{1,4}$, 
Jonathan~H. Newton$^{1,5}$, 
Aaron~C. Bell$^{1,6}$, 
Mary~E. Spraggs$^{1,4}$, 
J.~Marcus Hughes$^{1,4}$, 
Aaron~M. Tagliaboschi$^{1}$,                    
Christopher~M. Brunt$^{7}$, 
A.~Russell Taylor$^{8,9}$, 
Jeroen~M. Stil$^{9}$, 
\& 
Thomas~M. Dame$^{10}$
}

\affiliation{
$^1$Western Kentucky U., 
$^2$Union U., 
$^3$U. North Carolina-Chapel Hill, 
$^4$C. M. Gatton Acad.,
$^5$McMaster U., 
$^6$U. Tokyo, 
$^7$Exeter U., 
$^8$U. Calgary, 
$^9$U. Cape Town, 
$^{10}$Harvard-CfA
}

\pubyear{2015}
\volume{315}  
\jname{From Interstellar Clouds to Star-Forming Galaxies: Universal Processes?}
\editors{P. Jablonka, F. Van der Tak, \& P. Andr\'{e}, eds.}
\begin{document}

\maketitle

\begin{abstract}
We have mapped cold atomic gas in 21cm line {\sc H~i} self-absorption (HISA)
at arcminute resolution over more than 90\% of the Milky Way's disk.
To probe the formation of H$_2$ clouds, we 
have compared our HISA distribution with
CO $J\!=\!1\!-\!0$ line emission. 
Few HISA features in the outer Galaxy have CO at the same position
and velocity, while most inner-Galaxy HISA 
has overlapping CO.  But many apparent inner-Galaxy HISA-CO associations can be
explained as chance superpositions, so most inner-Galaxy HISA may also be
CO-free.  Since standard equilibrium cloud models cannot explain the very cold
{\sc H~i} in many HISA features without molecules being present, these clouds
may instead have significant CO-dark H$_2$.

\end{abstract}

\begin{figure}[t]
\begin{center}
 \includegraphics*[width=2.0in]{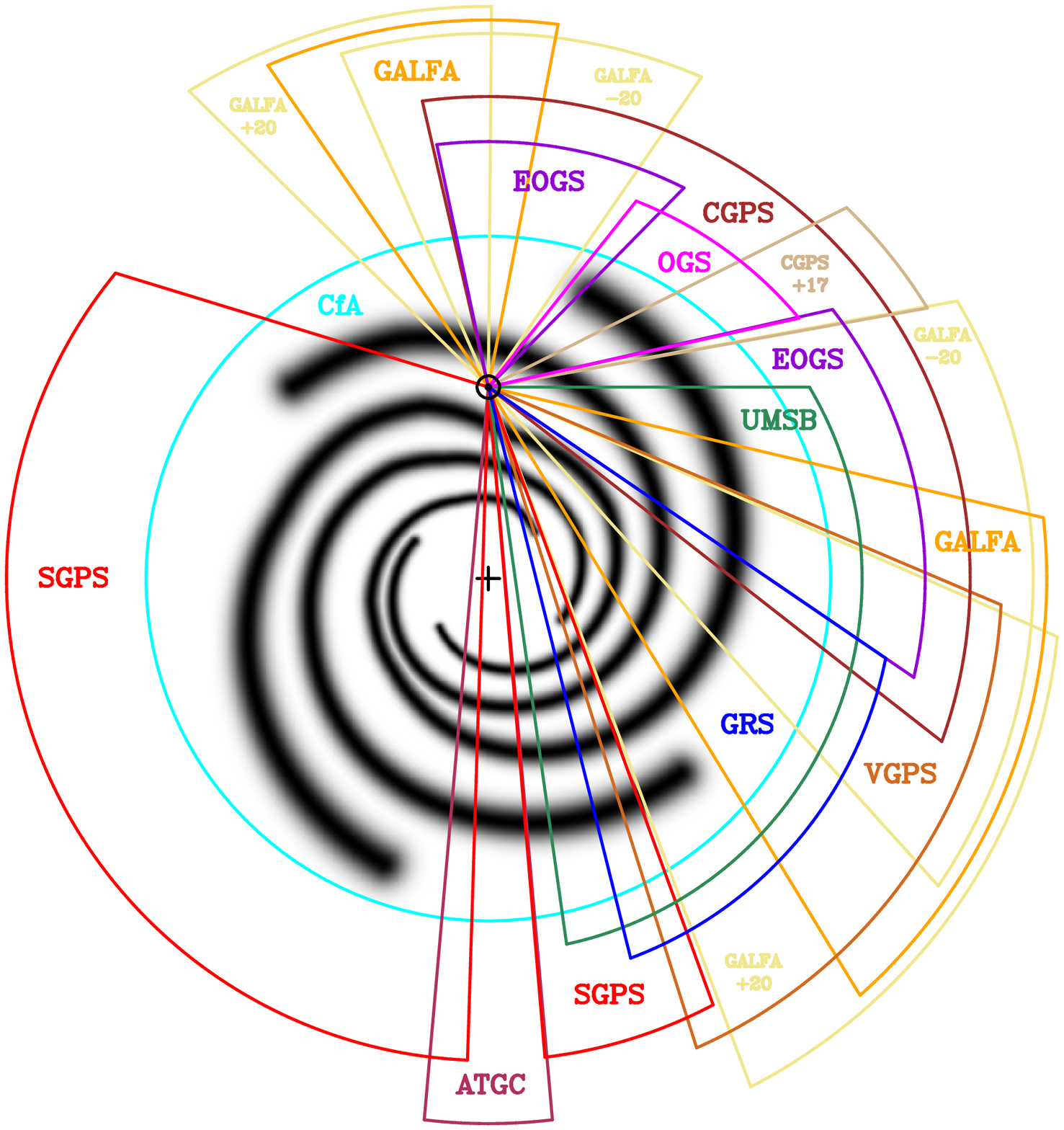} 
 \includegraphics*[width=3.0in,bb=50 330 562 676]{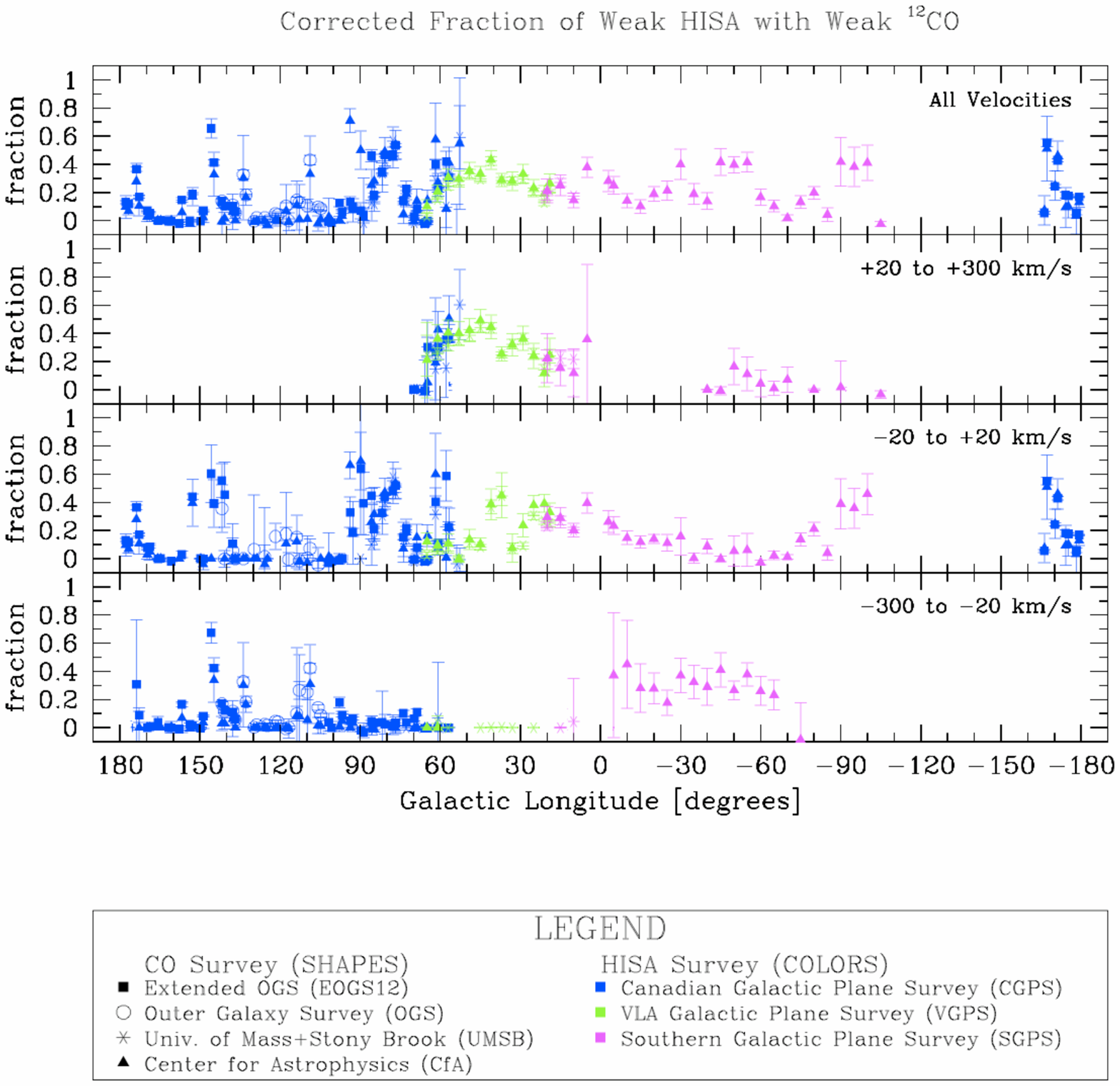}
 \caption{
{\it Left:\/}
Galactic disk coverage of {\sc H~i} and CO surveys used in this study
(see Table~1).  ATGC and GALFA data are not yet included.  The spiral arm model
is adapted from Taylor \& Cordes (1993).
{\it Right:\/} Corrected fractions of HISA voxels (volume pixels) with
$^{12}$CO emission at the same $(\ell,b,v)$ position, measured for different
survey data sets vs.\ longitude within different LSR velocity ranges to
separate trends for local, inner-Galaxy, and outer-Galaxy gas.  Points give the
mean fraction and $1\sigma$ error in the mean.  The plotted fraction values are
computed as the observed fraction minus the fraction expected if HISA and CO
are unrelated physically and only align by chance, where the latter fraction is
the product of the fraction of all voxels containing HISA and the fraction of
all voxels containing CO.  Threshholds used for significant detections are
$\Delta T_{b,{\rm HISA}} < -15$~K and $T_{b,{\rm 12CO}} > 1$~K.
}
   \label{fig1}
\end{center}
\end{figure}

\begin{table}
\centering
\caption{High-Resolution {\sc H~i} and CO Surveys Used in This Study}
\begin{tabular}{|c|c|c|c|c|c|}
 \hline
 Survey & Line & Telescope(s) & $\Delta\theta$ & $\Delta v$ & Plane Coverage, Area \\
 \hline
 CGPS$^1$ & {\sc H~i} 21cm & DRAO-ST + 26~m & 1$'$ & 0.8~km/s & $\!\!52^\circ\!<\!\ell\!<\!193^\circ$, 1240~deg$^2$$\!\!$ \\
 VGPS$^2$ & {\sc H~i} 21cm & VLA-D + GBT 100~m & 1$'$ & 0.8~km/s & $\!\!18^\circ\!<\!\ell\!<\!67^\circ$, 177~deg$^2$$\!\!$ \\
 SGPS$^3$ & {\sc H~i} 21cm & ATCA + Parkes 64~m & 2$'$ & 0.8~km/s & $\!\!253^\circ\!<\!\ell\!<\!20^\circ$, 274~deg$^2$$\!\!$ \\
 \hline
 OGS$^4$ & $^{12}$CO 1-0 & FCRAO~14~m & 1$'$ & 0.8~km/s & $103^\circ\!<\!\ell\!<\!142^\circ$, 328~deg$^2$ \\
 GRS$^5$ & $^{13}$CO 1-0 & FCRAO~14~m & 1$'$ & 0.2~km/s & $14^\circ\!<\!\ell\!<\!56^\circ$, 83~deg$^2$ \\
 EOGS$^6$ & $^{12+13}$CO & FCRAO~14~m & 1$'$ & 0.2~km/s & $56^\circ\!<\!\ell\!<\!192^\circ$, 820~deg$^2$ \\
 UMSB$^7$ & $^{12}$CO 1-0 & FCRAO~14~m & 6$'$ & 1.0~km/s & $8^\circ\!<\!\ell\!<\!90^\circ$, 164~deg$^2$ \\
 CfA$^{8}$ & $^{12}$CO 1-0 & CfA~1.2~m, N + S & 9$'$ & 0.6~km/s\ & $0^\circ\!<\!\ell\!<\!360^\circ$, 11,000~deg$^2$$\!\!$ \\
 \hline
    \multicolumn{6}{l}{\scriptsize
	$^{1}$Taylor et al.\ (2003);
	$^{2}$Stil et al.\ (2006);
	$^{3}$McClure-Griffiths et al.\ (2005);
	$^{4}$Heyer et al.\ (1998);
	} \\
    \multicolumn{6}{l}{\scriptsize
	$^{5}$Jackson et al.\ (2006);
	$^{6}$Brunt \etal\ (in prep);
	$^{7}$Sanders et al.\ (1986);
	$^{8}$Dame et al.\ (2001);
	} \\
\end{tabular}
\label{Tab:surveys}
\end{table}

\firstsection 
\section{Overview}

The gas in galactic disks occurs in a wide range of temperatures and densities,
most of which are unsuitable for star formation.  Somehow, diffuse atomic
clouds are collected into colder, denser molecular clouds that can collapse
under their own gravity.  Molecular condensation is not directly observable,
but it most likely arises in cold, quiescent pockets of atomic hydrogen ({\sc
  H~i}) gas, which over time will form molecular hydrogen (H$_2$) followed by
more observable molecular species.  Using algorithms developed previously
(Gibson \etal\ 2005a,b), we have mapped the cold {\sc H~i} population traced by
{\sc H~i} self-absorption (HISA; Gibson \etal\ 2000) against warmer {\sc H~i}
emission in three large {\sc H~i} synthesis surveys (Gibson 2010; see Table~1).
We then measured the mean fraction of positions with HISA that also have CO
emission in other surveys (Table~1; Figure~1) to evaluate the evolutionary
state of the HISA clouds.

We find that most HISA outside the Sun's orbit lacks CO emission at the same
position and velocity, while most inner-Galaxy HISA has overlapping CO, but the
latter may be illusory.  If the expected number of HISA-CO matches due to
chance alignments is removed, then the inner-Galaxy HISA-CO correspondence
drops below $\sim 50\%$.  Since HISA temperatures are too cold to explain
easily with purely atomic gas (Wolfire \etal\ 2003), many HISA features may
trace cold {\sc H~i} inside H$_2$ clouds that lack adequate UV shielding for
abundant CO (e.g., Wolfire \etal\ 2010).

We also find that CO positions with HISA are even less common than HISA
positions with CO, with a low enough fraction ($< 10\%$) to raise concerns
about the use of HISA to resolve near/far kinematic distance ambiguities in
inner-Galaxy sight lines.

Future steps in the analysis include corrections for FCRAO CO ``error beam''
sidelobe contamination, incorporation of other surveys, including GALFA-{\sc
  H~i} data from Arecibo, and comparison to synthetic observations of Galactic
disk models.

\vspace*{0.1in}
\noindent
{\bf Acknowledgements:} Support for this work was provided by the U.S. National
Science Foundation, NASA, Western Kentucky University, and the Gatton Academy.

\vspace*{0.25in}
\noindent
{\bf References}

\vspace*{0.1in}
{\normalfont\small
\noindent
\begin{minipage}[h]{0.45\textwidth}
{Dame, T. M. \etal} 2001, \textit{ApJ}, 547, 792 \\
{Gibson, S. J.} 2010, \textit{ASP-CS}, 438, 111  \\
{Gibson, S. J., \etal} 2005b, \textit{ApJ}, 626, 214 \\
{Gibson, S. J., \etal} 2005a, \textit{ApJ}, 626, 195 \\
{Gibson, S. J., \etal} 2000, \textit{ApJ}, 540, 851 \\
{Heyer, M. H., \etal} 1998, \textit{ApJS}, 115, 241 \\
{Jackson, J. M., \etal} 2006, \textit{ApJS}, 163, 145 \\
\end{minipage}
\hfill
\begin{minipage}[h]{0.45\textwidth}
{Sanders, D. B., \etal} 1986, \textit{ApJS}, 60, 1 \\
{Stil, J. M., \etal} 2006, \textit{AJ}, 132, 1158 \\
{Taylor, A. R., \etal} 2003, \textit{AJ}, 125, 3145 \\
{Taylor, J. H., \& Cordes, J. M.} 1993, \textit{ApJ}, 411, 674 \\
{Wolfire, M.~G., \etal} 2010, \textit{ApJ}, 716, 1191 \\
{Wolfire, M. G., \etal} 2003, \textit{ApJ}, 587, 278 \\
\end{minipage}

\vspace*{-0.1in}
\noindent
{McClure-Griffiths, N. M., \etal} 2005, \textit{ApJS}, 158, 178 
}
\end{document}